# Reducing COVID-19 Misinformation Spread by Introducing Information Diffusion Delay Using Agent-based Modeling

Mustafa Alassad, Nitin Agarwal

COSMOS Research Center, UA Little Rock, Little Rock, AR, USA
{mmalassad, nxagarwal}@ualr.edu

**Abstract**. With the explosive growth of the Coronavirus Pandemic (COVID-19), misinformation on social media has developed into a global phenomenon with widespread and detrimental societal effects. Despite recent progress and efforts in detecting COVID-19 misinformation on social media networks, this task remains challenging due to the complexity, diversity, multi-modality, and high costs of fact-checking or annotation. In this research, we introduce a systematic and multidisciplinary agent-based modeling approach to limit the spread of COVID-19 misinformation and interpret the dynamic actions of users and communities in evolutionary online (or offline) social media networks. Our model was applied to a Twitter network associated with an armed protest demonstration against the COVID-19 lockdown in Michigan state in May 2020. We implemented a one-median problem to categorize the Twitter network into six key communities (nodes) and identified information exchange (links) within the network. We measured the response time to COVID-19 misinformation spread in the network and employed a cybernetic organizational method to monitor the Twitter network. The overall misinformation mitigation strategy was evaluated, and agents were allocated to interact with the network based on the measured response time and feedback. The proposed model prioritized the communities based on the agents' response times at the operational level. It then optimized agent allocation to limit the spread of COVID-19 related misinformation from different communities, improved the information diffusion delay threshold to up to 3 minutes, and ultimately enhanced the mitigation process to reduce misinformation spread across the entire network.

## 1. Introduction

During the COVID-19 pandemic, many online social networks suffered from an "infodemic"—widespread circulation of misinformation or disinformation—that amplified the risks of the disease worldwide. Millions of online users shared information, stories, and experiences related to COVID-19 on social networks. Despite this fact, the officials in the World Health Organization (WHO) declared that COVID-19 related fake news is dangerous and spreading faster and more efficiently than the virus [1]. Likewise, President Biden criticized social media platforms for spreading misinformation about COVID-19 and upcoming vaccines on Friday, July 16, 2021 [2]. In addition, US officials raised alarms about a growing wave of misinformation about COVID-19 and related vaccines that threatened the administration's efforts to limit the pandemic [3].

Moreover, online users and communities reacted differently to COVID-19 related misinformation spread, where researchers said that at least 800 people may have died due to misinformation related to COVID-19 [4]. People worldwide witnessed much COVID-19 misinformation spread on social media networks connected to the safety guidelines, social distancing, mask-wearing, and lockdowns [5]. In some cases, these guidelines polarized communities, whereas some people behaved well by following the guidelines and applying the regulations for their safety. In some countries, a few communities did not accept the rules and started to react differently on social networks. They explained why they were concerned about these guidelines, while other groups escalated their reactions by protesting them; unfortunately, we witnessed armed protests against COVID-19 lockdowns [6].

Limiting the spread of misinformation in the context of COVID-19 is extremely important, where new false information may start from rumors and then spread across the social media network very fast. Misinformation could influence the understanding of many online users and, despite later proven false, can have harmful consequences for individuals and society [7]. Much effort was put into fighting COVID-19 misinformation spread on social networks.

WHO partnered with other parties, such as the UK, to provide healthier information on COVID-19 [5], and a few online tools and web applications are available to identify COVID-19 misinformation topics. For example, CoVerifi is a web application to verify COVID-19 related news [8]; the COSMOS research center developed a web application that tracks and educates about COVID-19 misinformation and provides tips on how to identify it [9]. In addition, many fact-checking tools, such as ClaimBuster [10], offer a near-complete fact-checking system. Google Fact Check [11] allows users to check claims, and GLTR (Giant Language model Test Room) classifies if the text is machine-generated or not [12]. Likewise, researchers implemented machine learning methods to detect fake news and COVID-19 related misinformation. Also, agent-based practices are used to study the dynamics of fear and behavior and their impact on combating the any other related abnormal actions from users and communities, as mentioned in [13], [14]. However, these methods' drawbacks concern counting and controlling the spread of COVID-19 misinformation between communities. Likewise, COVID-19 misinformation detection tools are developed to detect misinformation, but they are insufficient to stem and track the information traveling over social platforms. For example, Fact-checker tools need advanced methodologies and systematic approaches to understand the context, meaning of the contents, and intentions behind true versus false information spread [15].

Over the last few years, many other methods and models have been proposed to study the spread and limitation of misinformation between communities, individuals, and online/offline social networks. Authors [16] proposed an exploratory study into the propagation and content of misinformation on Twitter. They studied limiting the spread of misinformation in social networks investigated by [17], who studied when influential users adopt accurate information rather than false information campaigns on online social networks. Also, misinformation spreads on social media where the elements governing misinformation circulate between Facebook's users that consume information related to scientific and conspiracy news addressed [18]. However, modeling and optimizing the spread of misinformation is an NP-hard problem [19] due to the structures of the social networks, the behavior of the users and communities, information propagation across the social media networks, and the dynamicity of the reactions. Likewise, traditional graph theories such as the centrality and modularity methods fall short of identifying the focal information spreaders in online social media networks [20].

This research introduces a systematic agent-based model that simplifies the analysis and mitigates the misinformation spread on social networks. It simulates the behavior of online users and the regulations that can govern the mitigation of the COVID-19 related misinformation spread on social media networks. Similarly, this research allows for more straightforward methods to monitor the interactions between online users on social networks and the organizations willing to fight the misinformation spread on social media networks. For this purpose, we inherited the Systems Thinking and Modeling methods [21], [22] to simplify the complex behavior of users on social networks. Visualizing the actions of the agents mitigating the information spread, we built and evaluated the response strategy, enhanced the agents' activities, and performed a situational assessment of social network conditions to handle the misinformation spread over time. Furthermore, this multidisciplinary modeling was implemented to formulate, develop, and synthesize a set of optimal solutions to mitigate the misinformation spread in real-time. This modeling can respond to agents' operational needs in real-time and other dynamic constraints to simultaneously map the dynamic evolutions in the network.

Similarly, this modeling should select the optimal solutions that provide acceptable strategies to enhance the operational actions that respond to the misinformation spread on social media networks. Moreover, the modeling should optimize the selected results, assimilate the observable facts, and analyze misinformation's contributory causes and effects on social networks in real-time [21]. In addition, this modeling should simplify the complex analysis of misinformation spread in dynamic social networks and supplement agents with valuable tools that can diminish the nonlinearity and reduce the number of variables needed in the modeling.

Moreover, the agent-based model proposed here defines the assigned agents' behavior in response to misinformation spread and provides enhanced response strategy, optimizes the time response, and prioritizes the resources to track the spread of misinformation in dynamic networks over time. Furthermore, the presented model would enable the agents to monitor the critical communities and report the misinformation spread behavior between communities, record new misinformation spread (different sets of posts) on social media networks, and interact with the environment to limit the misinformation spread. We implemented the Organizational Cybernetic Approach (OCA) inherited from the Systems Thinking methods [22]. This method can control the communications between agents at the interaction level in any organization willing to fight misinformation spread and online/offline communities spreading COVID-19 related misinformation on social media networks. Also, the stochastic one-median problem [23], [24] is an operational method implemented in this model to enhance the agents' performance, find critical communities in the network, and enable the model to minimize the response time to any stochastic misinformation spread across the network.

This research makes several contributions to mitigating COVID-19 related misinformation spread. In the first contribution, we determine the upper bound (worst case scenario) of the misinformation spreaders in dynamic communities. The second contribution is to implement a procedure to evaluate the performance and availability of the agents interacting with online and offline social networks. The third contribution is to introduce a mathematical model to minimize the response time to COVID-19 misinformation spread on social networks and evaluate metrics that would reduce the agents' time to respond to any new misinformation spread in networks. The fourth contribution is that the model implements a multi-commodity network flow problem [25], [26] to simulate the misinformation spread between communities on social networks. The fifth contribution is that the model utilizes the information diffusion delay factor to enhance the agents' performance.

The rest of the paper is organized as follows. Section 2 reviews mitigating misinformation and agent-based modeling, systems thinking and modeling, and operational methods in social networks. Section 3 explains the solution strategy. Section 4 discusses the key outcomes, such as finding the best community combinations within the social network that will enhance the agents' monitoring process. It measures the agents' performance in handling the misinformation spread between communities, and it visualizes the agents' performance based on the feedback received from the online environment with and without adding the information diffusion delay factor up to a certain threshold level. Section 5 presents the conclusion and future work.

## 2. Related Work

This section reviews different topics related to the implemented agent-based methods, Systems Thinking and Modeling methods, and misinformation mitigation topics in social network analysis.

### 2.1. Misinformation Mitigation in Social Networks

The growth of misinformation on online social media hurts the entire society. It misleads people during health emergencies [27], especially during the COVID-19 pandemic. It has been challenging to track the traveling of COVID-19 misinformation across social networks [28] and to analyze the influenced users and communities over time [29]. Mitigating misinformation on social networks, mainly COVID-19 misinformation, has now been investigated by many scholars [8], [30], [31], ranging psychological studies [32], computational analysis [33], and many other perspectives. However, to limit the misinformation during the COVID-19 epidemic, the spread of misinformation was categorized into four categories: transmission, prevention, treatment, and vaccination [34]. Agent-based modeling was employed to study the dynamics of COVID-19 information spread [13], where Epstein's hypothesis of "coupled contagion dynamics of fear and disease" was used to explore fear-driven behavior adaptations and their impact on efforts to combat the COVID-19 pandemic. Similarly, the machine learning approaches were applied to study and detect the online fake news and misinformation related to the COVID-19 pandemic [35], where a multi-dimensional method was implemented by organizations and agents for detecting the fake news. Likewise, the social support, text mining, and media richness theories were employed to explore the different types of misinformation dissemination on social networks. The online rumor reversal model was introduced by [36]. The genetic algorithm-based rumor mitigation and precedence-based competitive cascade model propagated competing rumor and counter-rumor cascades in online social networks. These models were used to find the minimal set of seed users for the counter-rumor.

In addition, a multi-objective optimization problem was formulated to minimize the effect of rumors on the online social network [37]. According to [38], the machine learning methods used to detect the health misinformation spread in online communities are based on the elaboration likelihood model. In other related efforts, different methods were introduced to limit the misinformation spread on social networks, like the temporal influence blocking model presented in [39] and the time and budget constraint methods [40]. Observation-based policies are implemented to minimize the negative spread between communities in social networks [41]. Furthermore, using the point process-based intervention and reinforcement learning methods, the multistage intervention framework is used to tackle fake news in social networks [42]. Good versus bad campaigns on social networks were employed to limit the spread of misinformation [19]. In this research, an NP-hard greedy optimization problem was designed to address the influence limitation and identify a subset of individuals that need to be convinced to adopt accurate information to minimize the number of influenced users. In other research, two sub-problems were considered to mitigate misinformation in the social networks [43]. This research identifies the sources of misinformation and other influenced users in the network in the first sub-problem. The second sub-problem is to allocate a minimum number of monitors in the network to prevent misinformation campaigns.

### 2.2. Agent-based Modeling in Social Network

One of the main challenges when developing an agent-based model that responds to the spread of misinformation and fake news in social networks is to devise the simplest ways to simulate the nonlinear behavior of the users and feedback loops at the macro-level. Utilizing agent-based models has advantages in detecting COVID-19 related misinformation on social media, where such models would control the complexity of the analysis and the costs of fact-checking or annotation over time. In addition, agent-based models can control and scale up real-world experiments, and nonlinear dynamics can be introduced, a lack of which poses a significant problem in conventional experiments [44]. Also, agent-based models (ABM) were used to investigate the formation of opinions in interacting communities and whether polarization or separation emerged from this interaction [45]. Agent-based modeling can better explore the behavior of users- and large-scale communities and construct formal models, which require global behavior to verify the model against real-world phenomenon [46].

Furthermore, agent-based models study fear as an essential factor in users' behavior during an epidemic, as investigated in [46]. Finally, the research proposed in [47] used agent-based modeling to study the different behavioral outcomes from social theories, psychological theories, and empirical data that can be used to validate the model's results in the real world. The agent-based models were implemented in the social network [48], including different components that enable agents to make decisions, involve different environments, and adjust the rules and policies in real-time.

### 2.3. Systems Thinking and Modeling

Many empirical studies have pointed out that systems thinking and modeling are essential and the easiest ways to simplify complex systems [49][50]. For example, systems thinking methods like the Casual Feedback approach [49] investigate some of the variables needed in complex systems modeling and analysis. In addition, dynamic systems like social networks require the science of feedback, powerful simulations, and multiple loops in non-linear systems [51]. These methods would help managers understand the complex structure of networks, users' dynamic behavior, and communities' creation over time. Likewise, based on the received feedback from the system, professionals can implement new strategies concerning time and different states of equilibrium to enhance the system's functionality in the general [50]. Authors in [52] focused on implementing power laws in social networks and large-scale communities. Also, the authors of [53] studied systems thinking and modeling methods, like the boundary conditions approach, to describe information diffusion's temporal and spatial patterns over online social networks. The control theory [54] and the information theory [55] are used for modeling the information diffusion in social networks. Also, game theories and social networks approach [56] are used to enhance decision-makers' performance, and the authors stated that these combinations could bring optimized solutions and enable them to model different agent-based problems in social network analysis.

In summary, the uncertainties in users' behaviors and the stochasticity of the misinformation spread on online platforms increase the analysis complexity, making traditional methods fall short in modeling and tracking misinformation spread. To address this issue, we proposed a systematic agent-based methodology to enhance the communication between the online environment and the agent levels. This model is designed to mitigate COVID-19 misinformation in real-time and minimize the response time to any abnormal information spread. Additionally, it can implement a continuous monitoring procedure for the misinformation flow between online communities and users over time, enabling a more effective response to any changes in COVID-19 related misinformation spread in real-time.

## 3. Methodology

This section describes the overall research problem, the methods used to simplify the complex social network analysis, and the solution procedure. In addition, due to the formal models' requirements, we have included the assumptions that would help streamline the solution procedure. We have added systems thinking and modeling as complexity levels into our modeling to construct the agent-based modeling. This level was introduced as advanced classes of intelligence challenges combined progressively with the traditional community detection methods such as the modularity method [57] and the operation methods to mitigate the misinformation spread on social networks.

### 3.1. Problem Definition

Consider a social network $G = (N, A)$ consisting of the distinct node sets, $N = \{1,2, \ldots, n\}$ and the set of edges (links) $A = \{(i,j), (k,l), \ldots, (s,t)\}$ represented by directed node pair combinations going from community $i$ to community $j$. Communities $i$ and $j$ are associated with numerical values representing the number of intra edges $d_{j,i}$ between communities $i$ and $j$. These numbers represent the actual number of users from the community $i$ linked to users in community $j$ at every time window. Also, $h_j$ means the communities' rate of the misinformation spread in the worst-case scenario, and $\lambda_i$ is the proportion rate for an agent to handle the misinformation spread in the network.

The complexity in the problem statement is due to the complexity in the social media analysis; we design an agent-based model interacting with online social networks to mitigate COVID-19 misinformation spread, monitor the communities in real-time, report abnormal information spread in the environment, and respond to COVID-19 related misinformation spread.

### 3.2. Research Objectives

First, due to the excessive growth in the size of online social networks and complexities in the structure of online communities, initially the model used the one-median method to prioritize the communities in social networks, as shown in Figure 5. Second, mitigating the COVID-19 misinformation requires real-time and well-planned strategies. This research introduces an agent-based model that can interact with users and communities to mitigate the stochastic COVID-19 related misinformation spread in real-time, as shown in Figure 1 and Figure 2. Third, due to complexities in analyzing and the unpredicted behavior of online users, this research should be able to track down users' actions, measure agents' performance, and monitor the dynamic environment in real-time. The fourth objective is to utilize a feedback loop to measure and updates the behavior and the changes in the environment. Fifth, based on the feedback from the environment, the model should update the misinformation mitigation strategy and then inform the agent levels with new mitigation plans.

### 3.3. Systems Thinking and Modeling

This section describes the relationships between agents and online social networks and presents their connection as a complete social system. Modern approaches implemented by systems thinkers can investigate the interactions among local parts in systems and outside environments [51]. For this research, Systems Thinking and Modeling implements valuable tools to analyze complex social networks, trace the dynamic activities of online users and communities over time, and adjust agents' behavior based on the feedback loops from online social networks, as shown in Figure 1. This process would allow managers to change the misinformation mitigation strategies based on new community behavior or coordinating sets of information spreaders on social media networks and then push back a developed strategy into the environment. Likewise, this process would generate a complete system for misinformation mitigation on social networks. The following sections will cover the methods used to implement the model.

#### 3.3.1. Organizational Cybernetic Approach (OCA)

Organizational cybernetics is a system thinking and modeling method that studies and controls the communication between a system and its environments. This method includes a variety of negative and positive feedback loops, simultaneously recording changes in the behaviors of all parts of the system and the environment and witnessing the increase of complexities in the systems [51]. OCA should help the agents level monitor the online social networks. This method can perfectly optimize and inject the inputs into the environment concerning the dynamic changes (outputs) and feedback received from the online social network, as presented in Figure 1.

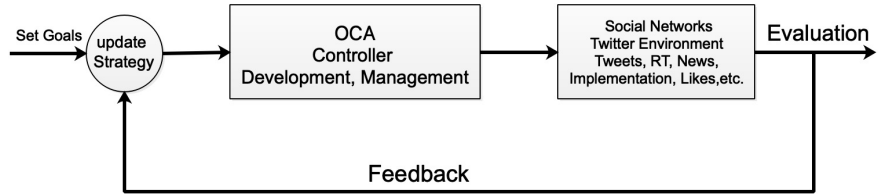

**Fig 1.** The social system with a feedback loop.

Moreover, the main aim is to limit COVID-19 related misinformation spread in online social networks, where the OCA is structured to communicate with the online environment as presented here:

OCA is structured to monitor, record feedback, and build strategies to enhance communication with the social environment, as shown in Figure 2:

- Develop and manage a plan to limit the misinformation spread,
- Assign agents to interact with the environment and monitor the sources of the misinformation spread,
- Evaluate the performance of the agents, and
- Develop a new strategy based on the environment's reported negative or positive feedback.

The OCA includes different levels to implement the mentioned goals, as presented in Figure 2, where the management level in OCA is needed to ensure fluency in the misinformation mitigation process. This level should send reports on 10 mitigation processes, agents' performance, and other related words to the control level; this level should speed the dynamic communication to the policy level in emergency scenarios. In addition, this level will help correct non-presented feedback and changes in the misinformation mitigation plans and lower any surprises in the social network that need to be immediately addressed.

The development level receives core information from the operational control level and information about changes in the online and offline environment. In other words, this level acts as an operating room in OCA to monitor the implementation of the misinformation mitigation process simultaneously at the agent level and in the environment.

The policy level ensures the system adapts to the misinformation and environment network as and when necessary to maintain the benefits gained from the mitigation process. This level should articulate the identity and purposes of the whole misinformation spread and reflect the entire network's strategy. However, the policy level can recruit experts to enhance the performance of the OCA and improve how different parts of the organization work as a whole and achieve their goals.

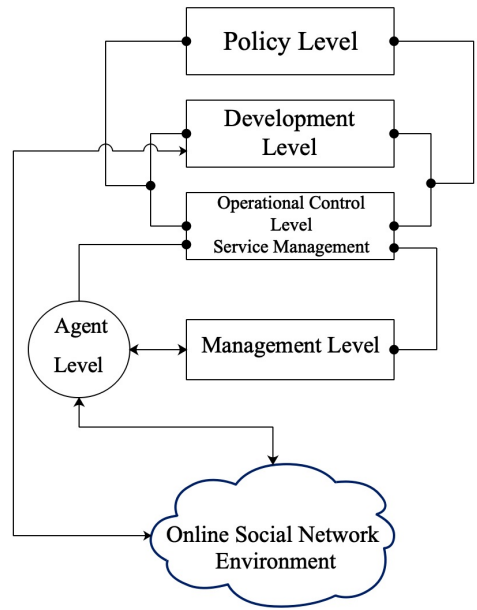

**Fig 2.** Organizational Cybernetic Approach.

Operation Level System (Agent level) in OCA is considered system #1 and needs to be as accessible as possible to allow its elements to interact with the outside environment efficiently [51]. The agent level has to maintain equity when monitoring the online network environment and considers factors such as first-in-first-out (FIFO) to report any abnormal behaviors or posts in the social networks to the management and to control levels both after and before completing or starting any reactions to the environment [58]. Moreover, this level may have the flexibility to deal with different communities or conduct other necessary actions in the network. Any agent at this level may interact with the network in various ways, such as blocking from the network the highly central rumors and influential accounts spreading COVID-19 misinformation.

### 3.4. Operation Research Method

To optimize the decisions and the actions of the agents whose work handles real-world problems in real-time, many operational methods are available that model the response to stochastic incidents and how the agents would minimize the resources to solve problems efficiently—for example, how to enhance the agents' performance in a fire department and their response to a highway car accident. How do they take the shortest path to travel? How can they get to the incident and clear the crash, while avoiding traffic during rush hour? Another example is how to enhance the agents' performance in response to any failures in telecommunication or power networks—how to restore the service fast to reduce network losses. To implement similar criteria for the agents working to mitigate the misinformation on social networks, in this section, we introduce the following methods to optimize the resources of the monitoring process and minimize the response time to COVID-19 misinformation spread on social networks. In this research, the agent level in OCA would not travel between communities. Still, the purpose was to help find and highlight the critical communities in a complex social network structure.

### 3.4.1. Stochastic One-Median Problem

The stochastic one-median problem is utilized in operation problems to study unpredictable incidents in dynamic networks, failures demanding fast operation response, and reactions to undetected issues in the real-time [59]. In other words, it is an allocation problem rather than a location problem. It can be helpful in social network analysis to identify critical communities based on the size and number of links with other communities in the structure of any social network. Furthermore, implementing this method should help prioritize the communities based on the agents' time response and use the best strategies rather than random reactions to the unpredicted behavior of users and communities on social networks over time.

The stochastic one-median problem has four outlines that can be translated into the structure of the OCA and the social network analysis, as presented below.
- The agent level is the sole system level in OCA responsible for response to abnormal behaviors in the social networks,
- The agent level would record and report abnormal interactions between the users and communities on the online social network,
- The agent level will report feedback and reactions to the higher levels in OCA, and
- This level must implement the off-scene setup time to respond to any stochastic and new abnormal behaviors on social networks.

The stochastic one-median problem developed from the following model. The objective function is to minimize the agents' expected response time to misinformation spread in the social environment, as shown in equation (1).

$$Min\ TR_j(C) \quad \forall j \in I \quad (1)$$

$$TR(C_j) = \bar{Q}_{C_j} + \bar{t}_{C_j} \quad \forall j \in I \quad (2)$$

$$\bar{Q}_{C_j} = \frac{\lambda_i\ \bar{S}_2(C_j)}{2\left(1 - \lambda_i \bar{S}(C_j)\right)} \quad \begin{array}{l}\forall j \in I \\ \forall i \in M\end{array} \quad (3)$$

$$\bar{t}_{C_j} = \sum_{j=1}^{I} h_j d(C_j, I) \quad \forall j \in I \quad (4)$$

$TR$ is the sum of the mean-queuing-delay $\bar{Q}$ and the mean response time $\bar{t}$ as shown in equation (2). Equation (3) defines $\bar{Q}$, where $C_j$ is the community $j$ in the network. $\lambda_i$ is the proportion rate for an agent to handle misinformation transferred between communities; for this purpose, the network is divided into sub-networks, as discussed in the next section. $\bar{S}(C_j)$ is the mean total response time (starting from the first moment the abnormal behavior is detected), and

$\bar{S}_2(C_j)$ is the second stochastic moment of the total response time to any new misinformation spread in the network. Equation (4) defines $\bar{t}$, where $I$ is the number of communities in the network, $h_j$ represents the rate of the misinformation spread from community $C_i$ in the worse scenario, and $d(C_j, I)$ is the shortest path to transfer information between community $C_j$ and community $C_i$.

### 3.4.2. Performance of the Agent Level

We present a linear multi-objective problem to evaluate the agents' performance in handling COVID-19 misinformation spread on social media. We model the information traveling between communities, measure the feedback from the environment, and finally apply the information diffusion delay factor to enhance the performance of the agents to mitigate the misinformation spread across the network. The multi-objective model includes three objectives. The first objective function Z1, is to calculate the actual average number of users that can spread misinformation between communities in the network. It is a network-reliability related problem and the users' connectivity in the community $i$ and community $j$ over time. The second objective function Z2, represents the shortest path; it is based on the minimum distance or a minimum number of nodes that the information can travel from users in community $i$ to users in community $j$. The advantage of this objective function is to prevent blocking and rerouting problems. The last objective function Z3, represents the information diffusion delay; this can illustrate the network reliability and the equity of the agents in the network. Through this objective function, each link from community $i$ to community $j$ is prevented from reaching its capacity. Likewise, the information diffusion delay tends toward infinity at the maximum possible connection between communities linked by the set of arcs U. These objective functions are associated with a set of constraints as follows:

$$Min\ Z1 = \sum_{p \in O} \sum_{q \in D} w^{pq} Y^{pq} \quad (5)$$
$$(information\ Spread)$$

$$Min\ Z2 = \sum_{u \in M} \sum_{p \in O} \sum_{q \in D} l_u X_u^{pq} \quad (6)$$
$$(Shortest\ Path)$$

$$Min\ Z3 = \sum_{u \in M} \left[ \frac{\sum_{p \in O} \sum_{q \in D} X_u^{pq}}{C_u - \sum_{p \in O} \sum_{q \in D} X_u^{pq}} \right] \quad (7)$$
$$(Data\ Delay)$$

Subject To:

$$\sum_{p \in O} \sum_{q \in D} X_u^{pq} \leq \frac{Delay \cdot \gamma \cdot \mu \cdot C_u}{|u|} \quad \forall u \in M \quad (8)$$

$$\sum_{u \in \Gamma(i)} X_u^{pq} - \sum_{u \in \Gamma^{-}(i)} X_u^{pq} = d^{pq} - Y^{pq} \quad if\ i = p \quad (9)$$

$$\sum_{u \in \Gamma(i)} X_u^{pq} - \sum_{u \in \Gamma^{-}(i)} X_u^{pq} = -d^{pq} + Y^{pq} \quad if\ i = q \quad (10)$$

$$\sum_{u \in \Gamma(i)} X_u^{pq} - \sum_{u \in \Gamma^{-}(i)} X_u^{pq} = 0 \quad otherwise \quad (11)$$

$$X_u^{pq}, Y^{pq} \geq 0, \forall u \in M, p \in O, q \in D$$

Equation 8 ensures that $X_u^{pq}$—the actual number of connections between two communities connected by arc $u$—is less than or equal to $C_u$ (the maximum possible connections between communities connected by arc $u$ when no delay is applied). $\gamma$ is the network throughput, similar to the total number of misinformation units in the network. $\mu = 1$ is a unit of information sent from a user in the community $i$ to a user in community $j$, and $Delay$ is the information diffusion delay factor. Equation 9 through Equation 11 are to measure the information unit conservation flow between communities in the network, where $d^{pq}$ is the number of monitored users between source community and destination community. $Y^{pq}$ is the number of not monitored users between source community and destination community. $\Gamma(i)$ is the set of arcs whose source community is $i$. $\Gamma-(i)$ is the set of arcs whose destination community is $i$. $w^{pq}$ is the amount of weight given to the communities $p$ and $q$. $l_u$ is the length of the path for information to transfer between communities. $p \in O$ is the source communities. $q \in D$ is the destination community setting. $N$ is the communities set in the network. $M$ is the arc (link) set for the network.

To summarize the operation research methodology, Figure 3 observes the solution procedure of the agent level, where OCA will assign the number of agent levels to monitor the social network. Graph theories were implemented to partition the network into smaller communities, as explained in section 4.1. Then, the stochastic one-median problem was implemented to prioritize the community selection, as described in section 3.4.1.

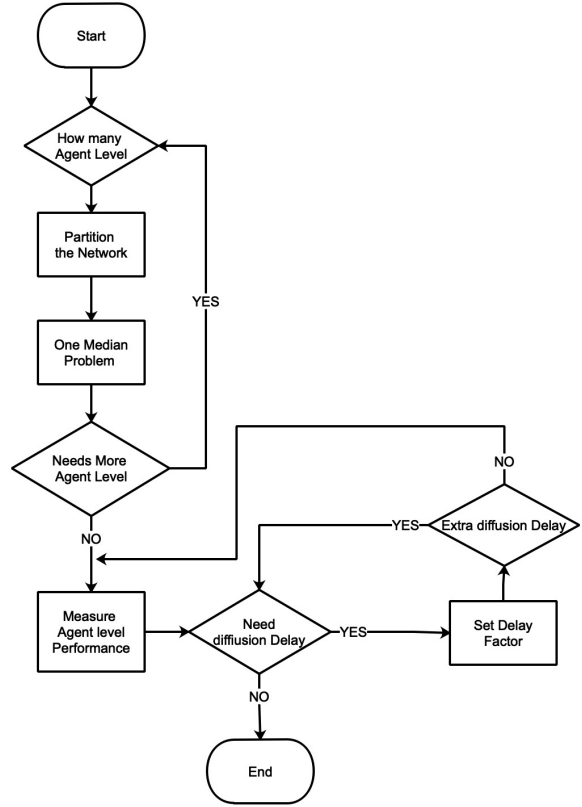

**Fig 3.** Flowchart for the solution procedure and performance evaluation of the agent level.

The next step is to measure the agent's performance in handling the misinformation spread in the network, considering the network throughput with or without the information diffusion delay variables as explained in section 3.4.2. In other words, based on the agents' performance and the feedback loop from the environment, the model should consider implementing a tuned diffusion delay factor to enhance the agents' performance in handling the misinformation spread up to a certain threshold level.

## 4.  Results

In this section, we present the data collection process on the Twitter network and the steps used to simplify the network's structure using graph theory methods. Next, we show the agent-based model solution procedure and evaluate the agents' performance in handling misinformation spread between communities. Finally, we implement the information diffusion delay to enhance the agent actions and discuss the outcomes.

### 4.1. Data Collection

This research used a Python script to collect a co-hashtag network using #MichiganProtest, #MiLeg, #Endthelockdown, and #LetMiPeopleGo hashtags from April 1st through May 20th, 2020 [30]. The behavior recorded polarized actions regarding COVID-19 related topics on Twitter, where some communities spread misinformation about the COVID-19 lockdowns in Michigan State. The data collected resulted in a 16,383 tweets network with 9,985 unique User IDs, as presented in Figure 4. This dataset revealed 3,632 nodes with 352 communities using the Modularity method [57]. The model focuses on the top five communities, as these were the communities that include the highest number of interactions between users. It is worth mentioning here that the model's findings do not depend on the data sample size; the only change that would occur if we added or removed more data for this model is the number of communities. As the number of communities changes, the interaction between the communities shown in Figure 5 changes. This means the possibilities of information exchanges increase or decrease based on the difference in the data sample. Other than that, the model that has been developed remains unaffected: the time it takes to run the model remains unaffected, the conclusion remains unaffected, and everything else remains unaffected after changing the data. The only change is that the graph in Figure 5 becomes either more or less connected. In addition, to simplify the network representation, the model represents all small communities in one node, accounting for any behavior or misinformation spread source from these small communities, in which the model will not change any linking behavior. We represent the small communities' links with the other five communities, as shown in Figure 5.

### 4.2. Assumptions

The following assumptions are considered to simplify the solution procedure:
- The policy level in OCA assigned only two agent levels to respond to the COVID-19 related misinformation spread on the Twitter network (M=2).
- Each agent level will handle less than ($\lambda_i$< 55%) of the entire network.
- The agent level response time to any misinformation spread is deterministic ($V = 55$) messages per hour.
- It is assumed that the maximum time required to respond to misinformation spread by any community is usually distributed ($\beta = 2$) case per hour.
- The time needed for the agent level to prepare for the second or the stochastic misinformation spread is deterministic, as shown in Appendix A.
- Communities are assumed to be active and do not disappear from the network over time.
- Communities are assumed to be spreading COVID-19 related misinformation.

Based on these assumptions, the steps explained in the next section and Appendix A show the solution procedure, prioritize the communities, and minimize the agents' response time.

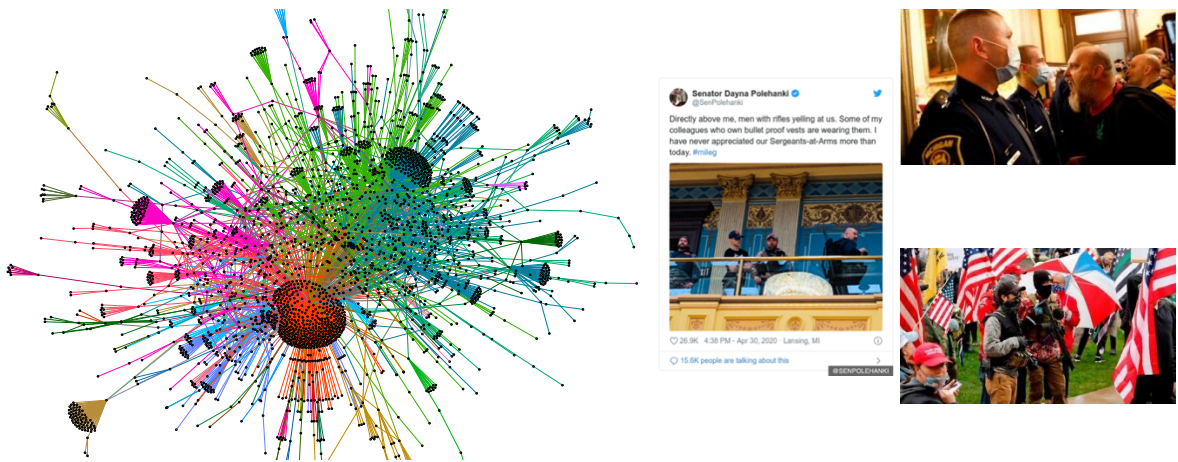

**Fig 4.** The network structure of the anti-COVID-19 lockdown in Michigan state.

## 4.3. Experimental Results

In this section, we explain the steps in the solution procedure, the model's overall structure, the process for evaluating the agents' performance, and the information diffusion delay factor.

Step 1: the network was clustered into smaller communities using the Modularity method [57] as presented in Figure 5, where this result identifies 352 communities as explained in section 3.4.1. We represent each community as $(C_i)$ where $I = \{1,2,...,6\}$, where $C_1$ through $C_5$ is the top five communities that have the highest number of users, and where $C_6$ represents all other communities.

Step 2: the intra edges between all six communities were measured; the links represent the $d(C_j, I)$ values as shown in Figure 5. These edges are the actual number of users in community $C_i$ linked to users in community $C_j$.

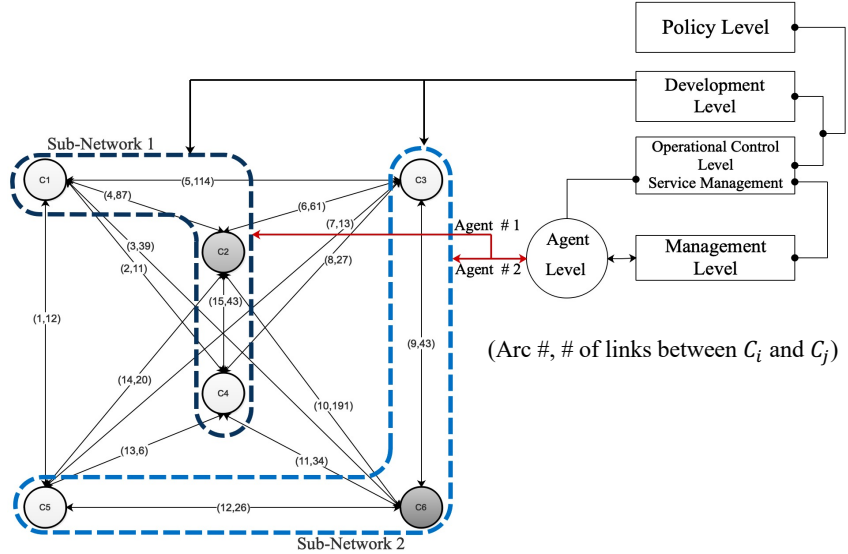

**Fig 5.** Overall model structure.

Step 3: the upper bound (worst case scenario) of the misinformation spreaders in communities $C_1$-$C_6$ was determined; Equations 12 and 13 represent the case when all users in $C_i$ spread misinformation in the network, indicating the communities' misinformation spread proportion rate $(h_i)$, where $I = \{1,2,...,6\}$ as shown in Table 1.

$$\sum_{j=1}^{I} h_j = 1 \quad (12)$$

$$h_j = \frac{|N_j|}{N} \quad (13)$$

Step 4: in this step, each agent is responsible for response to misinformation spread and monitoring less than 55% of the entire network. The network is divided into two sub-networks based on $\lambda_i$ values measured in Step 3 and shown in Table 1 and on the research assumption from section 4.2. Agent level #1 ($M_1$) is assigned to response to misinformation spread in sub-network 1, including $(C_1, C_2, C_4)$; Agent level #2 ($M_2$) is assigned to response to misinformation spread in sub-network 2, including $(C_3, C_5, C_6)$, as shown in Figure 5.

Step 5: the stochastic one-median model explained in section 3.4.1 was applied. The outcomes show the possible community combinations solutions for both agent levels to minimize the response time to misinformation spread. Table 2 presents the sorted nine possible solutions measured, where the best communities' combinations in the network for both agent levels to monitor were identified. In addition, based on the outcomes, the model assigned agent level #1 to monitor and respond to misinformation spread in $C_2$ and agent level #2 to respond to misinformation spread in $C_6$. The agents can minimize the response time and limit COVID-19 related misinformation spread when they first handle the spread in these communities.

**Table 1:** Misinformation spread proportion rate $h_j$ of Community $C_j$ in the network (worst case scenario).

| $h_{C_1}$ | $h_{C_2}$ | $h_{C_3}$ | $h_{C_4}$ | $h_{C_5}$ | $h_{C_6}$ |
|---|---|---|---|---|---|
| 0.2189 | 0.1944 | 0.1621 | 0.085 | 0.0656 | 0.274 |

These two communities include many users and are responsible for a high volume of information units exchanged from and to other communities in the network, as measured in step 2.

Nevertheless, the solution procedure ignored other communities that received a negative response time (infinity) due to their scales, locations, and the number of links they occupy in the network. Another reason is that simply monitoring these communities did not perform a fast response to the misinformation spread due to the heavy or small number of users in these communities, as we show in all the related calculations in Appendix A.

**Table 2:** Possible solutions combinations.

| $C_{i,j}$ | $Min\ TR$ (hours) |
|---|---|
| $C_2$ & $C_6$ | 7.699 & 0.618 |
| $C_2$ & $C_5$ | 7.699, & 1.399 |
| $C_2$ & $C_3$ | 7.699 & 1.944 |
| $C_1$ & $C_6$ | ∞ & 0.618 |
| $C_4$ & $C_6$ | ∞ & 0.618 |
| $C_1$ & $C_5$ | ∞ & 1.399 |
| $C_4$ & $C_5$ | ∞ & 1.399 |
| $C_1$ & $C_3$ | ∞ & 1.944 |
| $C_4$ & $C_3$ | ∞ & 1.944 |

**Table 3:** Reliability of network.

| Arc # | $C_i$ | $C_j$ | Reliability $a_{ij}$ | $X_{old}$ | $X_{new}$ |
|---|---|---|---|---|---|
| 1 | 1 | 5 | 65.3 | 12 | 8 |
| 2 | 1 | 4 | 68 | 11 | 8 |
| 3 | 1 | 6 | 65.3 | 39 | 25 |
| 4 | 1 | 2 | 68 | 87 | 59 |
| 5 | 1 | 3 | 65.3 | 114 | 74 |
| 6 | 3 | 2 | 65.3 | 61 | 40 |
| 7 | 3 | 5 | 97.4 | 13 | 13 |
| 8 | 3 | 4 | 65.3 | 27 | 18 |
| 9 | 3 | 6 | 97.4 | 43 | 42 |
| 10 | 2 | 6 | 65.3 | 191 | 125 |
| 11 | 4 | 6 | 65.3 | 13 | 8 |
| 12 | 5 | 6 | 97.4 | 26 | 25 |
| 13 | 4 | 5 | 65.3 | 6 | 4 |
| 14 | 2 | 5 | 65.3 | 13 | 13 |
| 15 | 2 | 4 | 68 | 43 | 29 |

### 4.4. Reliabilities of the Network for the Users-Users Links

In this section, we measure the reliability of the links between communities that spread misinformation, based on the response time measured in section 4.3. This step helps estimate and demonstrate the number of users who can spread misinformation between communities within 24-hour windows, as shown in Table 3 for all arcs U shown in Figure 5. Also, these numbers in Table 3 are based on sub-communities shown in Figure 5, where arcs (2, 4, and 15) lay in sub-network #1; this means that all misinformation spread on these arcs is handled by agent level #1. The same procedure applies to arcs (7, 9, and 12), where these communications are placed within sub-network #2 and where agent level #2 handles the misinformation spread for these parts of the network. However, other arcs (1,3,5,6,8,10,11,13,14) are in-between the two sub-networks, as shown in Figure 5; the reliability values are measured based on the combination of time response of both agent levels, as shown in Appendix B.

### 4.5. Evaluation of the Performance of the Agents and Feedback from the Twitter Environment

To simulate the performance of the agent levels to mitigate COVID-19 related misinformation spread in social media networks, we modeled the spread of misinformation between communities and up to 250 misinformation units spread between communities in the network, as shown in Figure 6. The network's throughput is the total number of misinformation units spread, as shown in Table 4. Each misinformation unit includes sets of messages transferred from the community $C_i$ (source) to community $C_j$ (target). In other words, the model will measure the agents' performance in mitigating up to five different misinformation spread units between communities in the network. Also, Figure 6 shows the misinformation units spread from users in $C_1$ (source) to users in $C_2$ and $C_4$ (targets), where $C_1$ will send 100 misinformation units, 50 per each target community. Likewise, 50 misinformation units spread from $C_2$ to communities in $C_6$. Another misinformation spread was recorded from $C_5$ to $C_3$. Finally, 50 misinformation units spread from communities in $C_6$ to target users in community $C_2$.

Table 4 shows the number of misinformation units sent from source communities to the target communities, with the number of misinformation units agents mitigated or handled and the number of misinformation units that were spread between communities (or not handled) in the network.

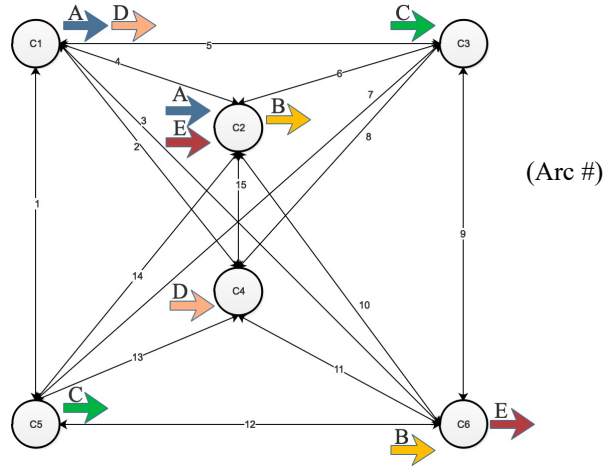

**Fig 6.** Misinformation spread between communities.

In other words, agents in both levels could monitor the networks, record the misinformation spread, and respond to more the 97% of the misinformation spread between communities in the network. These results show the evaluation step after implementing the mitigation strategy as shown in Figure 1; the feedback loop from the environment that would deliver the performance of agent level #2 failed to handle seven misinformation units sent from node $C_6$ to users in community $C_2$. The following section discusses the agents' performance and how to enhance the misinformation mitigation procedure.

Based on the measured performance from agents shown in Table 4, the model recorded seven unhandled misinformation units spread between users in node $C_6$ and users in node $C_2$. To adjust the mitigation strategy and enhance the agent levels' performance, we employed the information diffusion delay factor in the solution procedure in this section. Then we measured the agents' performance and compare outcomes, which helped measure the agents' reaction concerning the diffusion delay and the number of actual users who could spread information over time.

The information diffusion delay was applied to the equation (7) shown below from section 3.4.2, where the data delay thresholds of (0.01–0.1) per hour were used to evaluate the agents' performance in mitigating the misinformation spread in the network as shown in Table 5. The information diffusion delay factor directly impacted the links between communities (the actual number of users linked between communities). Increasing the delay threshold allowed the agent to handle more users and misinformation units spread between communities. Furthermore, the information diffusion delay factor depends on the total number of users able to spread misinformation units across the network (network throughput), the number of links available in the network (actual number of users linked between communities), and the average message unit length transferred between communities in the network as shown below.

Table 5 shows the actual number of users in the network agents can handle after adding the delay factors into the model's solution procedure. The model presumes that the agents should handle misinformation units when the information diffusion delay is added to the model. Figure 7 shows the efficient frontier after implementing the information diffusion delay at the selected thresholds. These values indicate that the effect of the delay factor does not start to impact the agents' performance to mitigate the misinformation spread until the delay threshold is set to less than 0.05 hours. Also, it shows a considerable delay effect when it is less than 0.01 hours.

$$\sum_{youthe\ p \in O} \sum_{q \in D} X_u^{p.q} = \frac{Delay \cdot \gamma \cdot \mu \cdot C_u}{Link}$$

$$Link = 15$$
$$\gamma = 250, total\ network\ throughput$$
$$\mu = 1, actual\ average\ number\ of\ misinformation\ units.$$

Table 4: Agent-levels performance to mitigate the misinformation spread between communities in real-time.

| Info. units | $C_i$ | $C_j$ | # of misinformation units | Handled | Not handled |
|---|---|---|---|---|---|
| A | 1 | 2 | 50 | 50 | 0 |
| B | 2 | 6 | 50 | 50 | 0 |
| C | 5 | 3 | 50 | 50 | 0 |
| D | 1 | 4 | 50 | 50 | 0 |
| E | 6 | 2 | 50 | 43 | 7 |
| Total | | | 250 | 243 | 7 |

In addition, implementing the information diffusion delay optimizes agents' resources and mitigates the spread of misinformation when the delay factor is set to 0.05 or higher. This strategy enhances the performance of agents to handle the maximum number of misinformation units spread between different communities on social networks. Appendix C shows the implemented code in Lingo.

Table 5: Information diffusion delay factor.

| Arc # | $C_i$ | $C_j$ | Information Diffusion Delay Thresholds | | | | | | |
|---|---|---|---|---|---|---|---|---|---|
| | | | None | 0.01 | 0.02 | 0.03 | 0.04 | 0.05 | 0.1 |
| 1 | 1 | 5 | 8 | 1 | 2 | 4 | 5 | 7 | 13 |
| 2 | 1 | 4 | 8 | 1 | 2 | 4 | 5 | 7 | 13 |
| 3 | 1 | 6 | 25 | 4 | 8 | 12 | 17 | 21 | 42 |
| 4 | 1 | 2 | 59 | 10 | 19 | 29 | 38 | 50 | 98 |
| 5 | 1 | 3 | 74 | 12 | 24 | 37 | 48 | 62 | 123 |
| 6 | 3 | 2 | 40 | 7 | 13 | 20 | 26 | 33 | 67 |
| 7 | 3 | 5 | 13 | 2 | 4 | 6 | 8 | 11 | 22 |
| 8 | 3 | 4 | 18 | 3 | 5 | 9 | 11 | 15 | 30 |
| 9 | 3 | 6 | 42 | 7 | 13 | 21 | 27 | 35 | 70 |
| 10 | 2 | 6 | 125 | 21 | 41 | 62 | 82 | 104 | 208 |
| 11 | 4 | 6 | 8 | 1 | 2 | 4 | 5 | 7 | 13 |
| 12 | 5 | 6 | 25 | 4 | 8 | 12 | 17 | 21 | 42 |
| 13 | 4 | 5 | 4 | 0 | 1 | 2 | 2 | 3 | 7 |
| 14 | 2 | 5 | 13 | 2 | 4 | 6 | 8 | 11 | 22 |
| 15 | 2 | 4 | 29 | 9 | 9 | 14 | 19 | 24 | 48 |

## 4.6. Discussion

The main findings and technical and practical implications are discussed in this section.

### 4.6.1. Main Findings

Mitigating the spread of misinformation on social media networks is a complicated process that needs advanced tools and new methods to overcome various existing problems. First, the implementation of the agent-based model is a helpful tool to interact with online and offline environments to monitor the behavior of users in real-time. Second, depending on the network's structure, the delay factor methodology can improve the misinformation mitigation process on social media networks. Third, the systems thinking methods are the right path to implementing advanced social networks analysis tools. Fourth, the results enhanced the mitigation misinformation diffusion process where the optimized threshold was 0.05 hour (3 min) and there were two agent levels to mitigate the misinformation completely for our case study.

### 4.6.2. Theoretical and Practical Implications

This research's theoretical and practical implications are presented in implementing a systematic method to mitigate the COVID-19 related misinformation spread on social networks. First, this multidisciplinary model is novel because it integrates the operation methods and the organization cybernetic methods to reduce the misinformation spread (front-end and back-end practical methods). Second, the model can control the real-time communications between the agents and the online environment. Third, the model can monitor the changes in the communities' behavior, measure the agents' performance, and adjust the mitigation strategies in real-time. Fourth, the model can handle complex social network analysis regardless of the size and number of network users. Fifth, the model proposed optimized methods to track the traveling of the misinformation, considering the community's behavior and COVID-19 misinformation diffusion mitigation process in real-time with the minimum number of resources (two agent levels) for this case study. Sixth, this model enhanced the decisions of the managers to optimize their resource allocation based on the size of the social network, the number of communities, and the feedback from the environment in real-time.

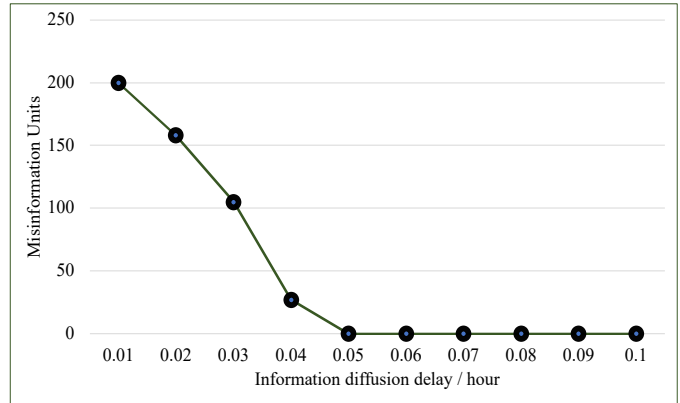

**Fig 7.** Performance of the agents in handling the misinformation spread after adding the information diffusion delay factors.

## 5. Conclusion and Future Directions

The recent challenges for mitigating COVID-19 related misinformation spread in social networks project the necessity for using advanced multidisciplinary approaches such as systems thinking and operational methods in social network analysis. In this research, we encountered challenges in forming a novel systematic model able to simplify the complexities in the structure of social networks, implement strategies to mitigate misinformation, and maximize the model's performance concerning the changes in the environment and feedback from real-world social networks.

An agent-based model was implemented to mitigate the spread of COVID-19 related misinformation in social networks; the Organization Cybernetic Approach was implemented to manage the mitigation process and allow the agents at the operation level to communicate with the environment easily. Also, the one-median problem was employed to enhance the agents' response. This method involved optimizing the resources, understanding the complex structure of the social networks, and improving the effectiveness of the agents. Consequently, the agents' performance was evaluated when the model considered the information diffusion delay. Finally, the results illustrated that the agents could completely handle and mitigate the misinformation spread in the Twitter network when the delay factor was set up to a certain threshold.

For future work, we will study and measure the increase and decrease of misinformation spread in dynamic social networks. In addition, we will evaluate the performance of the OCA's strategies to mitigate the misinformation spread after considering different approaches like reinforcement learning and training maze-agent.

## Acknowledgment

This research is funded in part by the U.S. National Science Foundation (OIA-1946391, OIA-1920920, IIS-1636933, ACI- 1429160, and IIS-1110868), U.S. Office of the Under Secretary of Defense for Research and Engineering (FA9550-22-1-0332), U.S. Army Research Office (W911NF-20-1-0262, W911NF-16-1-0189, W911NF-23-1-0011, W911NF-24-1-0078), U.S. Office of Naval Research (N00014-10-1-0091, N00014-14-1-0489, N00014-15-P- 1187, N00014-16-1-2016, N00014-16-1-2412, N00014-17-1-2675, N00014-17-1-2605, N68335-19-C-0359, N00014-19-1-2336, N68335-20-C-0540, N00014-21-1-2121, N00014-21-1-2765, N00014-22-1-2318), U.S. Air Force Research Laboratory, U.S. Defense Advanced Research Projects Agency (W31P4Q-17-C- 0059), Arkansas Research Alliance, the Jerry L. Maulden/Entergy Endowment at the University of Arkansas at Little Rock, and the Australian Department of Defense Strategic Policy Grants Program (SPGP) (award number: 2020-106-094). Any opinions, findings, and conclusions or recommendations expressed in this material are those of the authors and do not necessarily reflect the views of the funding organizations. The researchers gratefully acknowledge the support.

**Appendix A**

This step is associated with two assumptions where ($V=55$, and $\beta =2$)

**Sub-Network 1: includes communities $C_1$, $C_2$, $C_4$.**

**Community # 1 ($C_1$)**

$$\bar{t}_{c_1} = \frac{h_{C_2}}{\lambda_1} \cdot \frac{d_{C_2,C_1}}{V} + \frac{h_{C_4}}{\lambda_1} \cdot \frac{d_{C_2,C_1} + d_{C_2,C_4}}{V} \qquad \bar{t}_{c_1} = 1.020$$

$$\bar{S}_{c_1} = \beta \cdot \bar{t}_{c_1} \qquad \bar{S}_{c_1} = 2.041$$

$$\bar{S}_{2_{c_1}} = \frac{h_{C_2}}{\lambda_1} \cdot \left[\frac{\beta \cdot (d_{C_2,C_1})}{V}\right]^2 + \frac{h_{C_4}}{\lambda_1} \cdot \left[\frac{\beta \cdot (d_{C_2,C_1} + d_{C_2,C_4})}{V}\right]^2 \qquad \bar{S}_{2_{c_1}} = 7.717$$

$$TR(C_1) = \frac{\lambda_1 \bar{S}_{2_{c_1}}}{2(1 - \lambda_1 \bar{S}_{c_1})} + \bar{t}_{c_1} \qquad TR(C_1) = -113.222$$

**Community # 2 ($C_2$)**

$$\bar{t}_{c_2} = \frac{h_{C_1}}{\lambda_1} \cdot \frac{d_{C_2,C_1}}{V} + \frac{h_{C_4}}{\lambda_1} \cdot \frac{d_{C_2,C_4}}{V} \qquad \bar{t}_{c_2} = 0.828$$

$$\bar{S}_{c_2} = \beta \cdot \bar{t}_{c_2} \qquad \bar{S}_{c_2} = 1.656$$

$$\bar{S}_{2_{c_2}} = \frac{h_{C_1}}{\lambda_1} \cdot \left[\frac{\beta \cdot (d_{C_2,C_1})}{V}\right]^2 + \frac{h_{C_4}}{\lambda_1} \cdot \left[\frac{\beta \cdot (d_{C_2,C_4})}{V}\right]^2 \qquad \bar{S}_{2_{c_2}} = 4.814$$

$$TR(C_2) = \frac{\lambda_1 \bar{S}_{2_{c_2}}}{2(1 - \lambda_1 \bar{S}_{c_2})} + \bar{t}_{c_2} \qquad TR(C_2) = 7.699$$

**Community # 4 ($C_4$)**

$$\bar{t}_{c_4} = \frac{h_{C_2}}{\lambda_1} \cdot \frac{d_{C_2,C_4}}{V} + \frac{h_{C_1}}{\lambda_1} \cdot \frac{d_{C_2,C_4} + d_{C_2,C_1}}{V} \qquad \bar{t}_{c_4} = 1.343$$

$$\bar{S}_{c_4} = \beta \cdot \bar{t}_{c_4} \qquad \bar{S}_{c_4} = 2.041$$

$$\bar{S}_{2_{c_4}} = \frac{h_{C_2}}{\lambda_1} \cdot \left[\frac{\beta \cdot (d_{C_2,C_4})}{V}\right]^2 + \frac{h_{C_1}}{\lambda_1} \cdot \left[\frac{\beta \cdot (d_{C_2,C_4} + d_{C_2,C_1})}{V}\right]^2 \qquad \bar{S}_{2_{c_4}} = 7.717$$

$$TR(C_4) = \frac{\lambda_1 \bar{S}_{2_{c_4}}}{2(1 - \lambda_1 \bar{S}_{c_4})} + \bar{t}_{c_4} \qquad TR(C_4) = -6.578$$

**Sub-Network 2: includes communities $C_3$, $C_5$, $C_6$.**

**Community # 3 ($C_3$)**

$$\bar{t}_{c_3} = \frac{h_{C_6}}{\lambda_2} \cdot \frac{d_{C_6,C_3}}{V} + \frac{h_{C_5}}{\lambda_2} \cdot \frac{d_{C_6,C_3} + d_{C_6,C_5}}{V} \qquad \bar{t}_{c_3} = 0.553$$

$$\bar{S}_{c_3} = \beta \cdot \bar{t}_{c_3} \qquad \bar{S}_{c_3} = 1.107$$

$$\bar{S}_{2_{c_3}} = \frac{h_{C_6}}{\lambda_2} \cdot \left[\frac{\beta \cdot (d_{C_6,C_3})}{V}\right]^2 + \frac{h_{C_5}}{\lambda_2} \cdot \left[\frac{\beta \cdot (d_{C_6,C_3} + d_{C_6,C_5})}{V}\right]^2 \qquad \bar{S}_{2_{c_3}} = 1.963$$

$$TR(C_3) = \frac{\lambda_2 \, \bar{S}_{2_{c_3}}}{2(1 - \lambda_2 \bar{S}_{c_3})} + \bar{t}_{c_3} \qquad TR(C_3) = 1.944$$

**Community # 6 ($C_6$)**

$$\bar{t}_{c_6} = \frac{h_{c_3}}{\lambda_2} \cdot \frac{d_{c_6,c_3}}{V} + \frac{h_{c_5}}{\lambda_2} \cdot \frac{d_{c_6,c_5}}{V} \qquad \bar{t}_{c_6} = 0.286$$

$$\bar{S}_{c_6} = \beta \cdot \bar{t}_{c_6} \qquad \bar{S}_{c_6} = 0.572$$

$$\bar{S}_{2_{c_6}} = \frac{h_{c_3}}{\lambda_2} \cdot \left[\frac{\beta \cdot (d_{c_2,c_1})}{V}\right]^2 + \frac{h_{c_5}}{\lambda_2} \cdot \left[\frac{\beta \cdot (d_{c_6,c_5})}{V}\right]^2 \qquad \bar{S}_{2_{c_6}} = 0.825$$

$$TR(C_6) = \frac{\lambda_2 \, \bar{S}_{2_{c_6}}}{2(1 - \lambda_2 \bar{S}_{c_6})} + \bar{t}_{c_6} \qquad TR(C_6) = 0.618$$

**Community # 5 ($C_5$)**

$$\bar{t}_{c_5} = \frac{h_{c_5}}{\lambda_2} \cdot \frac{d_{c_6,c_5}}{V} + \frac{h_{c_3}}{\lambda_2} \cdot \frac{d_{c_6,c_5} + d_{c_6,c_3}}{V} \qquad \bar{t}_{c_5} = 0.425$$

$$\bar{S}_{c_5} = \beta \cdot \bar{t}_{c_5} \qquad \bar{S}_{c_5} = 0.850$$

$$\bar{S}_{2_{c_5}} = \frac{h_{c_5}}{\lambda_2} \cdot \left[\frac{\beta \cdot (d_{c_6,c_5})}{V}\right]^2 + \frac{h_{c_3}}{\lambda_2} \cdot \left[\frac{\beta \cdot (d_{c_6,c_5} + d_{c_6,c_3})}{V}\right]^2 \qquad \bar{S}_{2_{c_5}} = 1.876$$

$$TR(C_5) = \frac{\lambda_2 \, \bar{S}_{2_{c_5}}}{2(1 - \lambda_5 \bar{S}_{c_5})} + \bar{t}_{c_5} \qquad TR(C_5) = 1.399$$

**Appendix B**

| | |
|---|---|
| $R_{Arc\,2_{4,15}} = \left(1 - \frac{TR_{C_2}}{24}\right) * 100 = 68\%$ | $X_2 = \left(\frac{68}{100}\right) * 11 = 8$ |
| $R_{Arc\,7_{9,12}} = \left(1 - \frac{TR_{C_6}}{24}\right) * 100 = 97.4\%$ | $X_7 = \left(\frac{97.4}{100}\right) * 13 = 13$ |
| $R_{Arc\,1,3,5\atop 6,8,10,11,\,13,14} = \left(1 - \frac{TR_{C_2} + TR_{C_6}}{24}\right) * 100 = 65.3\%$ | $X_1 = \left(\frac{65.3}{100}\right) * 12 = 8$ |

**Appendix C**

```
SETS:
    NODE ;      ! set of nodes;
    COMMUNITY;  ! set of Communities;
  Delay;
    ARC(NODE, NODE):  ! Some of the node pairs have  links;
     U,    !Total capacity;
     FLO; !Flow on this arc, determined;
    CXN( COMMUNITY, NODE): ! External Flow _Combinations of Communities
and nodes;
     B1, !Demand - supply at this node of each Community;
   Y1,
  Y2;  ! Lost Call;
    CXA( COMMUNITY, ARC):
     C,  ! Cost/unit flow Community k on this arc;
     UK, ! Upper bound on flow  of Community k on this arc;
   LK, ! lower bound on flow of Community k on this arc;
```

```
     X;  ! Flow of Community k on this arc;
 DXA (Delay) :
 D;
 ENDSETS
 DATA:
  COMMUNITY = 1..5;
  NODE = 1..6;
   ! Demand - supply at each node;
   b1 =  -50  50 0  0  0  0      ! Community A;
          0   -50  0  0  0  50   ! Community B;
          0   0   50 0  -50 0    ! Community C;
          -50 0   0  50 0   0    ! Community D;
          0   50  0  0  0   -50; ! Community E;

   ! Arcs available, and capacities over all;
    ARC   U =
   1 2 59
   1 3 74
   1 4 8
   1 5 8
   1 6 25
   2 1 59
   2 3 40
   2 4 29
   2 5 13
   2 6 125
   3 1 74
   3 2 40
   3 4 18
   3 5 13
   3 6 42
   4 1 8
   4 2 29
   4 3 18
   4 5 4
   4 6 8
   5 1 8
   5 2 13
   5 3 13
   5 4 4
   5 6 25
   6 1 25
   6 2 125
   6 3 42
   6 4 8
   6 5 25
;

    ! Community, Arc capacities and costs;

   UK   C =
  !1 2;  59 1 ! Community 1;
   !1 3;  74 1
   !1 4;  8  1
   !1 5;  8  1
   !1 6;  25   1
   !2 1;  59   1
   !2 3;  40   1
   !2 4;  29   1
```

```
!2 5;   13    1
!2 6;   125   1
!3 1;   74    1
!3 2;   40    1
!3 4;   18    1
!3 5;   13    1
!3 6;   42    1
!4 1;   8 1
!4 2;   29    1
!4 3;   18    1
!4 5;   4 1
!4 6;   8 1
!5 1;   8 1
!5 2;   13    1
!5 3;   13    1
!5 4;   4 1
!5 6;   25    1
!6 1;   25    1
!6 2;   125   1
!6 3;   42    1
!6 4;   8 1
!6 5;   25    1

!1 2;   59 1  ! Community 2;
 !1 3;  74 1
 !1 4;  8 1
 !1 5;  8 1
 !1 6;  25    1
!2 1;   59    1
!2 3;   40    1
!2 4;   29    1
!2 5;   13    1
!2 6;   125   1
!3 1;   74    1
!3 2;   40    1
!3 4;   18    1
!3 5;   13    1
!3 6;   42    1
!4 1;   8 1
!4 2;   29    1
!4 3;   18    1
!4 5;   4 1
!4 6;   8 1
!5 1;   8 1
!5 2;   13    1
!5 3;   13    1
!5 4;   4 1
!5 6;   25    1
!6 1;   25    1
!6 2;   125   1
!6 3;   42    1
!6 4;   8 1
!6 5;   25    1

!1 2;   59 1  ! Community 3;
 !1 3;  74 1
 !1 4;  8 1
 !1 5;  8 1
 !1 6;  25    1
!2 1;   59    1
```

```
!2 3;   40    1
!2 4;   29    1
!2 5;   13    1
!2 6;   125   1
!3 1;   74    1
!3 2;   40    1
!3 4;   18    1
!3 5;   13    1
!3 6;   42    1
!4 1;   8 1
!4 2;   29    1
!4 3;   18    1
!4 5;   4 1
!4 6;   8 1
!5 1;   8 1
!5 2;   13    1
!5 3;   13    1
!5 4;   4 1
!5 6;   25    1
!6 1;   25    1
!6 2;   125   1
!6 3;   42    1
!6 4;   8 1
!6 5;   25    1

!1 2;   59 1  ! Community 4;
 !1 3;  74 1
 !1 4;  8 1
 !1 5;  8 1
 !1 6;  25    1
!2 1;   59    1
!2 3;   40    1
!2 4;   29    1
!2 5;   13    1
!2 6;   125   1
!3 1;   74    1
!3 2;   40    1
!3 4;   18    1
!3 5;   13    1
!3 6;   42    1
!4 1;   8 1
!4 2;   29    1
!4 3;   18    1
!4 5;   4 1
!4 6;   8 1
!5 1;   8 1
!5 2;   13    1
!5 3;   13    1
!5 4;   4 1
!5 6;   25    1
!6 1;   25    1
!6 2;   125   1
!6 3;   42    1
!6 4;   8 1
!6 5;   25    1

!1 2;   59 1  ! Community 5;
 !1 3;  74 1
 !1 4;  8 1
 !1 5;  8 1
```

```
    !1 6;   25    1
    !2 1;   59    1
    !2 3;   40    1
    !2 4;   29    1
    !2 5;   13    1
    !2 6;   125   1
    !3 1;   74    1
    !3 2;   40    1
    !3 4;   18    1
    !3 5;   13    1
    !3 6;   42    1
    !4 1;   8  1
    !4 2;   29    1
    !4 3;   18    1
    !4 5;   4  1
    !4 6;   8  1
    !5 1;   8  1
    !5 2;   13    1
    !5 3;   13    1
    !5 4;   4  1
    !5 6;   25    1
    !6 1;   25    1
    !6 2;   125   1
    !6 3;   42    1
    !6 4;   8  1
    !6 5;   25    1

  ;
    ENDDATA

    ! MINIMIZE Cost of flow;
     MIN=   ( @SUM(CXA( k,i,j): X(k,i,j)*C(k,i,j)) + 100*(@SUM(CXN(k,i):
Y1(k,i)+Y2(k,i)))   );

    ! Conservation of flow for each Community k  at node i:
      Flow in - Flow out >= demand (- supply);
      @FOR( CXN( k, i):
        @SUM(  ARC(  j,i):  X(k,j,i))   -  @SUM(ARC(i,j):  X(k,i,j))   =
B1(k,i)+Y1(k,i)-Y2(k,i);
         );

    ! Capacity on arc i,j over all Communities;
      @FOR( ARC(i,j):
         FLO(i,j) = @SUM( COMMUNITY(k): X(k,i,j));
       FLO(j,i) = @SUM( COMMUNITY(k): X(k,j,i));
         (FLO(i,j) + FLO(j,i)) <= U(i,j)*0.33;
         );

    ! Capacity on flow of Community k on arc i,j;
      @FOR( CXA(k,i,j):
         X(k,i,j) <= UK(k,i,j);

         );
```